\documentclass{llncs}
\usepackage{url}
\usepackage{epsfig}

\begin{document}

\title{MIPS code compression}
\author{Tommi Hirvola\inst{1}}
\institute{Aalto University, Finland\\
    \email{tommi.hirvola@aalto.fi}}
\maketitle

\begin{abstract}

MIPS machine code is very structured: registers used before are likely to be
used again, some instructions and registers are used more heavily than others,
some instructions often follow each other and so on. Standard file compression
utilities, such as gzip and bzip2, does not take full advantage of the
structure because they work on byte-boundaries and don't see the underlying
instruction fields. My idea is to filter opcodes, registers and immediates from
MIPS binary code into distinct streams and compress them individually to
achieve better compression ratios. Several different ways to split MIPS code
into streams are considered. The results presented in this paper shows that a
simple filter can reduce final compressed size by up to 10~\% with gzip and
bzip2.

\end{abstract}

\section{Introduction}

MIPS is a RISC instruction set architecture that is mostly used in embedded
devices. MIPS, like many other RISC architectures, uses 4-byte fixed-length
instructions to simplify hardware design, but this in turn decreases the code
density compared to CISC architectures. Embedded devices have limited amount of
memory and it is often the limiting factor on what the system can do. Better
compression can also reduce the production costs of embedded systems. In this
paper I aim to preprocess MIPS code for better compressibility by splitting
machine code instructions into separate streams with high autocorrelation.

Similar techniques have been applied to x86 program code with a great success.
\emph{PPMexe}~\cite{PPM} by Drinic et al preprocesses x86 code by reordering
instructions and splitting the program binary into sub-streams that have high
autocorrelation. According to their results, the preprocessing combined with an
improved version of PPM algorithm reduced final compression size by 18--24~\%
with the binaries used. \emph{kkrunchy}~\cite{kkr} by Giesen is a
state-of-the-art compressor for 64K demoscene intros at the time of writing.
kkrunchy splits x86 instructions into separate streams based on their function,
and according to Giesen this improves compression ratios by about 10~\% with
LZ-based and context-based coders. Many EXE compressors, including Microsoft's
Cabinet (CAB) archive format and 7-Zip, utilizes simple filters to transfrom
relative call offsets into absolute addresses, and this alone can improve
compression ratio by 10~\% for x86 code. The relative-to-absolute offset
transform is not applicable for MIPS, because MIPS already uses absolute
addresses to encode function calls.

There has been previous work done on MIPS code compression as well. MIPS
Technologies introduced MIPS16e code-compression extension to MIPS architecures
in 2004. MIPS16e increases the code density by including a second 16-bit
instruction set that covers a subset of commonly used MIPS instructions.
According to MIPS Technologies, applications compiled with MIPS16e are 30~\%
smaller on average and suffer a small (15~\%) performance hit. Lekatsas and
Wolf~\cite{CCES} have proposed a dictionary-based algorithm for MIPS code
compression. Their algorithm, referred to as \emph{SADC} (Semiadaptive
Dictionary Compression), replaces commonly used opcode-operand combinations
with extended 8-bit opcodes and splits the remaining instructions into
independent streams. SADC uses 4 streams: opcodes, registers, 16-bit immediates
and 32-bit immediates. In Section~\ref{sec:experiments}, I show that this
particular instruction subdivision doesn't work well with gzip or bzip2.

I will be focusing on MIPS R3000 processor which has the following three general
instruction formats:

\begin{figure}
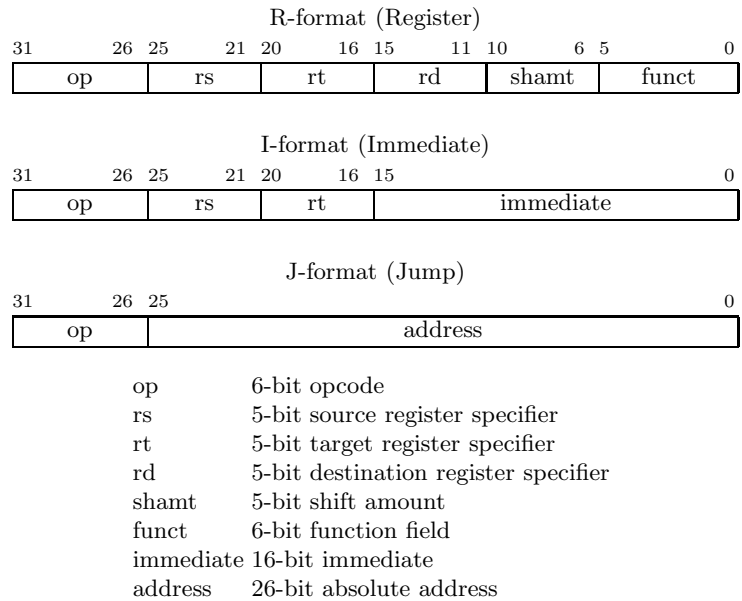

  \begin{center}
    \newcommand{\bitsr}[2]{{\scriptsize #1 \hfill #2 \hspace{1mm}}}
    \newcommand{\bitsrr}[2]{{\scriptsize #1 \hfill #2}}
    \newcommand{\fldc}[1]{\multicolumn{1}{|c}{#1}}
    \newcommand{\fldcc}[1]{\multicolumn{1}{|c|}{#1}}
    R-format (Register) \\
    \begin{tabular}{@{}p{1.8cm} @{}p{1.5cm} @{}p{1.5cm} @{}p{1.5cm} @{}p{1.5cm} @{}p{1.8cm}}
      \bitsr{31}{26} & \bitsr{25}{21} & \bitsr{20}{16} & \bitsr{15}{11} & \bitsr{10}{6} & \bitsrr{5}{0} \\
      \hline
      \fldc{op} & \fldc{rs} & \fldc{rt} & \fldc{rd} & \fldc{shamt} & \fldcc{funct} \\
      \hline
    \end{tabular} \\ [0.5cm]

    I-format (Immediate) \\
    \begin{tabular}{@{}p{1.8cm} @{}p{1.5cm} @{}p{1.5cm} @{}p{4.8cm}}
      \bitsr{31}{26} & \bitsr{25}{21} & \bitsr{20}{16} & \bitsrr{15}{0} \\
      \hline
      \fldc{op} & \fldc{rs} & \fldc{rt} & \fldcc{immediate} \\
      \hline
    \end{tabular} \\ [0.5cm]

    J-format (Jump) \\
    \begin{tabular}{@{}p{1.8cm} @{}p{7.8cm}}
      \bitsr{31}{26} & \bitsrr{25}{0} \\
      \hline
      \fldc{op} & \fldcc{address} \\
      \hline
    \end{tabular} \\ [0.3cm]

    {\footnotesize
    \begin{tabular}{l l}
      op & 6-bit opcode \\
      rs & 5-bit source register specifier \\
      rt & 5-bit target register specifier \\
      rd & 5-bit destination register specifier \\
      shamt & 5-bit shift amount \\
      funct & 6-bit function field \\
      immediate & 16-bit immediate \\
      address & 26-bit absolute address
    \end{tabular} }
    \caption{MIPS instruction formats}
    \label{fig:formats}
  \end{center}
\end{figure}

\section{Split-stream filter}

The goal of filtering is to split instructions into separate streams with high
similarity in values in the same stream, but little similarity between values
in different streams. The streams then can be compressed and decompressed with
existing algorithms and tools. This process is called \emph{split-stream
encoding} \cite{PPM}\cite{kkr}. The filtering is a reversible process.

Choosing the right streams (i.e.\ what instruction fields to store together) is
a critical part, because it is what determines the compressibility of the
streams. One way would be to store all instruction fields into separate
streams, but this gives poor results for several reasons. For example, register
specifier fields are 5-bits width and storing them into a distinct stream
breaks byte-alignment of that stream, which drastically hurts compression of
character-based algorithms such as gzip and bzip2 (at least for small files).
This problem can be avoided by choosing the stream subdivision carefully and
using padding if needed. Also, we will lose any correlation between registers
and opcodes if we filter them into separate streams (some registers and
instructions are often used together, e.g.\ \verb!jal $ra!). Splitting the
instructions into too many small streams has a negative impact impact on the
compression ratio~\cite{PPM}. Several different ways of dividing code into
streams are discussed and experimented in Section~\ref{sec:experiments}.

Implementing the split-stream filter requires a rough disassembler engine. The
(dis)assembler engine used in my implementation determines the instruction type
(R, I, J) from an opcode and allows easy manipulation of instruction fields.
The disassembler also recognizes three types of 16-bit immediates so that these
can be separated: branch offsets, load/store offsets and constants. The filter
disassembles instructions passed to it and splits them into streams according to
predetermined rules. The filtering begins at the first instruction of the code
and proceeds to the last instuction. 

Unfiltering (\emph{merging}) works as follows: an opcode is read from a
predefined opcode stream, which tells us the instruction format and what
instruction fields needs to be read next. Then the instruction field values are
read from the corresponding streams and assembled into one instruction. This
process is repeated until there is no more opcodes left in the opcode stream.
Note that the opcode stream may also contain other data besides opcodes (e.g.\
registers/immediates), but this does not cause problems as long as everything
is written and read in a correct order.

The filter must preserve invalid and unrecognizable code. Code sections often
contain jump tables and other data that isn't decodable as MIPS instructions,
and corrupting such data is not acceptable. In my implementation unknown
instructions are decoded similarly to J-format instructions: the unknown opcode
is stored in the opcode stream and the remaining 26-bits are stored into a
stream called \emph{unknown}.

The filtering and unfiltering algorithms are very fast, and can be easily
implemented in less than 200 lines of C code when targeting a specific
processor.

\section{Experiments} \label{sec:experiments}

\begin{figure}[!ht]
  \begin{center}
    \epsfig{file=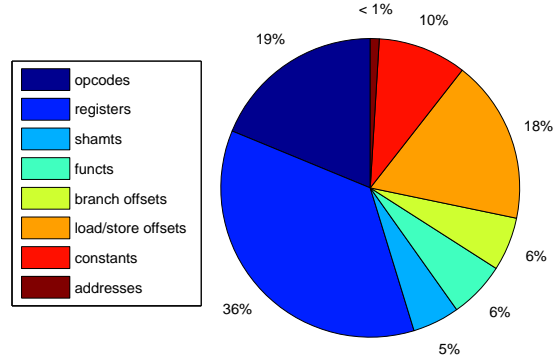, height=2.0in, clip=}
  \end{center}
  \caption{Instruction field distribution of all.text in terms of size. Registers (rs, rt,
           rd) are combined into one section and 16-bit immediates are divided
           into three sections: branch offsets, load/store offsets and
           constants.}
  \label{fig:rij}
\end{figure}

Text segments of \verb!/bin/! UNIX utilities of debian-mips (2.6.21-5-4kc-malta)
were used as a test data. Files that had text segment smaller than 50~KB were
discarded. The remaining 25 files were concatenated to create one large test
file: \verb!all.text!. In addition, Python and GIMP were compiled to produce two
more big test files. See Figure~\ref{fig:final} for details of these files.
Figure~\ref{fig:rij} shows the size distribution of instruction fields of
\verb!all.text!.

\begin{figure}[!ht]
  \begin{center}
    \epsfig{file=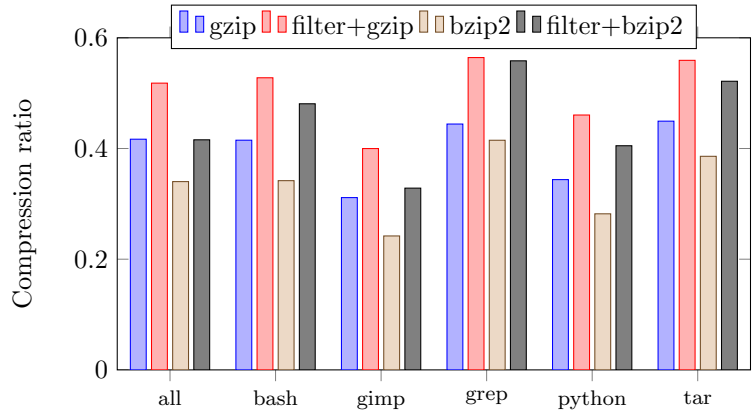}
    \caption{7 streams: opcodes, registers, functs, shamts, imm16, imm26 \& unknown}
    \label{fig:compr0}
  \end{center}
\end{figure}

My initial idea was to use 7 streams: opcodes, registers, functs, shamts, 16-bit
immediates, 26-bit immediates and unknown data. This turned out to work poorly,
as shown in Figure~\ref{fig:compr0}.

Compression ratios got 20--45~\% worse with the filter in all test cases. There
are multiple reasons for this: the streams are not byte-aligned (except imm16),
some streams are very small, correlations between registers and opcodes are
lost etc. All the exact reasons are not clear. Padding values to byte boundary
before storing them into the streams improved compression ratios by about 20~\%
compared to no padding at all, but the filtering still had no positive effect.

\begin{figure}[!ht]
  \begin{center}
    \epsfig{file=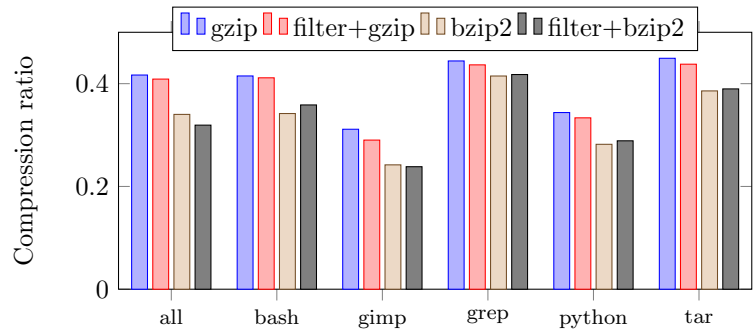}
    \caption{2 streams: core, 16-bit immediates}
    \label{fig:compr1}
  \end{center}
\end{figure}

The next idea was to have two streams: one for 16-bit immediates and the other
for everything else (\emph{core stream}). The intuition is that 16-bit
immediates don't have any correlation with the core stream, and therefore
separating them only reduces ``context dilution'' and should not harm the
compression. This stream subdivision also preserves the byte-alignment, because
only aligned 16-bit values are removed from the original data. The results are
shown in Figure~\ref{fig:compr1}.

Now, the filtering improved gzip compression by around 5~\% in all test cases.
bzip2's compression ratios of \verb!all! and \verb!gimp! were improved by 6~\%
and 1~\%, respectively. In all test cases the compression ratio of the core
stream was 10--20~\% better than the compression ratio of the ``unfiltered''
file, but the 16-bit immediates were compressed poorly which increased the
overall compression ratio.

\begin{figure}[!ht]
  \begin{center}
    \epsfig{file=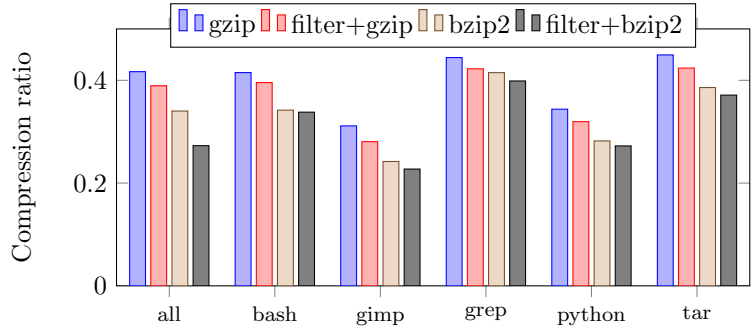}
    \caption{4 streams: core, branch offsets, load/store offsets \& 16-bit constants}
    \label{fig:compr2}
  \end{center}
\end{figure}

The next step was to make the immediates to compress better. MIPS has three
types of 16-bit immediates: branch offsets, load/store offsets and constants.
Intuitively, the immediates differ from each other significantly. Most of the
branch instructions jump back and forth in one function, and therefore have
offset value near to zero, whereas the load/store offsets are often used to
access structure members. So, the next step was to divide 16-bit immediates
into three streams: branch offsets, load/store offsets and 16-bit contants. The
results are shown in Figure~\ref{fig:compr2}.

\newpage

Using these four streams, the filter improved compression ratio of all the test
files --- including 22 test files not shown in the figure. The biggest
improvement (20~\% with bzip2) was in \verb!all.text! file, which was created
by concatenating all test files that were larger than 50~KB (excluding
\verb!gimp! and \verb!python!). This might be because bzip2 loses context
between files and the filter somehow mitigates the effect (?). The detailed
compression ratios are presented in Figure~\ref{fig:final}.

\begin{figure}[!ht]
  \begin{center}
    {\footnotesize
      \begin{tabular}{|l|r||r|r|r||r|r|r|}
        \hline
        File        &     size &     gzip & filter+gzip & impr. &    bzip2 & filter+bzip2 & impr. \\
        \hline
        all.text    & 4 825 KB & 2 011 KB &    1 878 KB &  7 \% & 1 642 KB &     1 315 KB & 20 \% \\
        gimp.text   & 4 126 KB & 1 285 KB &    1 158 KB & 10 \% &   999 KB &       938 KB &  6 \% \\
        python.text & 1 657 KB &   570 KB &      530 KB &  7 \% &   468 KB &       451 KB &  3 \% \\
        bash.text   &   785 KB &   326 KB &      311 KB &  5 \% &   269 KB &       266 KB &  1 \% \\
        tar.text    &   344 KB &   155 KB &      146 KB &  6 \% &   133 KB &       128 KB &  4 \% \\
        grep.text   &   112 KB &    50 KB &       48 KB &  5 \% &    47 KB &        45 KB &  4 \% \\
        \hline
      \end{tabular}}
    \caption{Final results using 4 streams: core, branch, load/store and
             constants. Uncompressed size, compressed size with/without filter
             and improvement are shown in the table.}
  \label{fig:final}
  \end{center}
\end{figure}

Several other ways to split instructions into streams were also tested. Even
though registers take up a lot of space (see Figure~\ref{fig:rij}), filtering
them into a separate stream doesn't seem beneficial (with or without padding).
Separating functs and shamts from the core stream improved bzip2's compression
by a few tenths of a percent more, but made gzip perform worse.

\newpage
\section{Conclusions and future work}

The results show that splitting MIPS instructions into streams can reduce
compressed sizes by up to 10~\% (and up to 20~\% in some special cases). An
interesting result was that the ``obvious'' subdivision of streams had no
positive effect at all (Figure~\ref{fig:compr0}). The best way that I found to
divide MIPS code into streams was to separate only 16-bit immediates from the
instructions, and then divide them into branch offsets, load/store offsets and
constants. This kind of filter is very fast and easy to implement.

The filter can be combined with other transforms to increase compressibility of
MIPS binaries even further. PPMexe\cite{PPM} uses a split-stream filter with
instruction scheduling algorithm to make x86 code more compressible, and the
same approach can be used with MIPS. Replacing commonly used instruction with
short symbols (see MIPS16e and SDAC~\cite{CCES}) is also one option, but it
might not work well with the split-stream filter. In my implementation the
streams are stored into different files. One stream may provide an adequate
context for compressing other streams, and in that case concatenating the
streams could increase the compressibility even more.

In this paper I focused on gzip and bzip2, other algorithms may benefit more or
less from the split-stream filter. It would be interesting to see how the
filter and different stream subdivisions affect other algorithms. Using a
word-based algorithm that is not constrained to use byte-aligned symbols could
have a big impact on the compression ratio.

\bibliographystyle{plain}
\bibliography{mips}

\end{document}